\documentclass[final]{ias2}

\usepackage{graphicx} 
\usepackage{multirow}
\usepackage{array} 

\usepackage{hyperref} 
 \usepackage{refcheck} 
\def\SUSEFLAV{{\tt SuSeFLAV}}

\def\LAPACK{{\tt LAPACK}}

\def\MICROMEGAS{{\tt MICROMEGAS}}
\def\DARKSUSY{{\tt DarkSUSY}}
\def\SUPERISO{{\tt SuperIso}}

\def\beq{\begin{equation}}
\def\eeq{\end{equation}}

\begin{document}

\markboth{SuSeFLAV:  A program for calculating supersymmetric spectra and lepton flavor violation}{D Chowdhury et. al.,}

\title{SuSeFLAV:  A program for calculating supersymmetric spectra and lepton flavor violation}

\author[sin]{Debtosh Chowdhury} 
\email{debtosh@cts.iisc.ernet.in}
\author[sin]{Raghuveer Garani} 
\email{rgarani@cts.iisc.ernet.in}
\author[sin]{Sudhir K. Vempati}
\email{vempati@cts.iisc.ernet.in}
\address[sin]{Centre For High Energy Physics, Indian Institute of Science, Bangalore 560012, India}

\begin{abstract}
We introduce the program \SUSEFLAV ~ for computing supersymmetric mass spectra with flavor violation in various
supersymmetric breaking scenarios with/without seesaw mechanism.  A short user guide summarizing the 
compilation, executables and the input files is provided.  
\end{abstract}

\keywords{MSSM, Right Handed Neutrinos, Lepton Flavor Violation}

\pacs{11.30.Pb, 12.15.Ff, 12.60.Jv, 13.35.Bv, 13.35.Dx, 14.60.Pq, 14.60.St.}
 
\maketitle


\textbf{1.}~Presently,  low energy supersymmetry is facing its stiffest challenge in terms of direct 
experimental searches at LHC and indirect searches through flavor violating rare processes and dark matter experiments. 
It is thus important to have a complete program which takes into account  flavor violation in the Yukawa
couplings as well as in the soft sector.  The presence of flavor violation through seesaw or otherwise
 can significantly modify the regions of parameter space which are compatible with experimental results. 
 \SUSEFLAV \cite{suseflavdoc}~ is designed to consider flavor violation both the Yukawa couplings as well as mass matrix
 computations. 
The short summary of the program is given below. 

\begin{figure}[ht]
\begin{center}
\includegraphics[width=0.9\columnwidth]{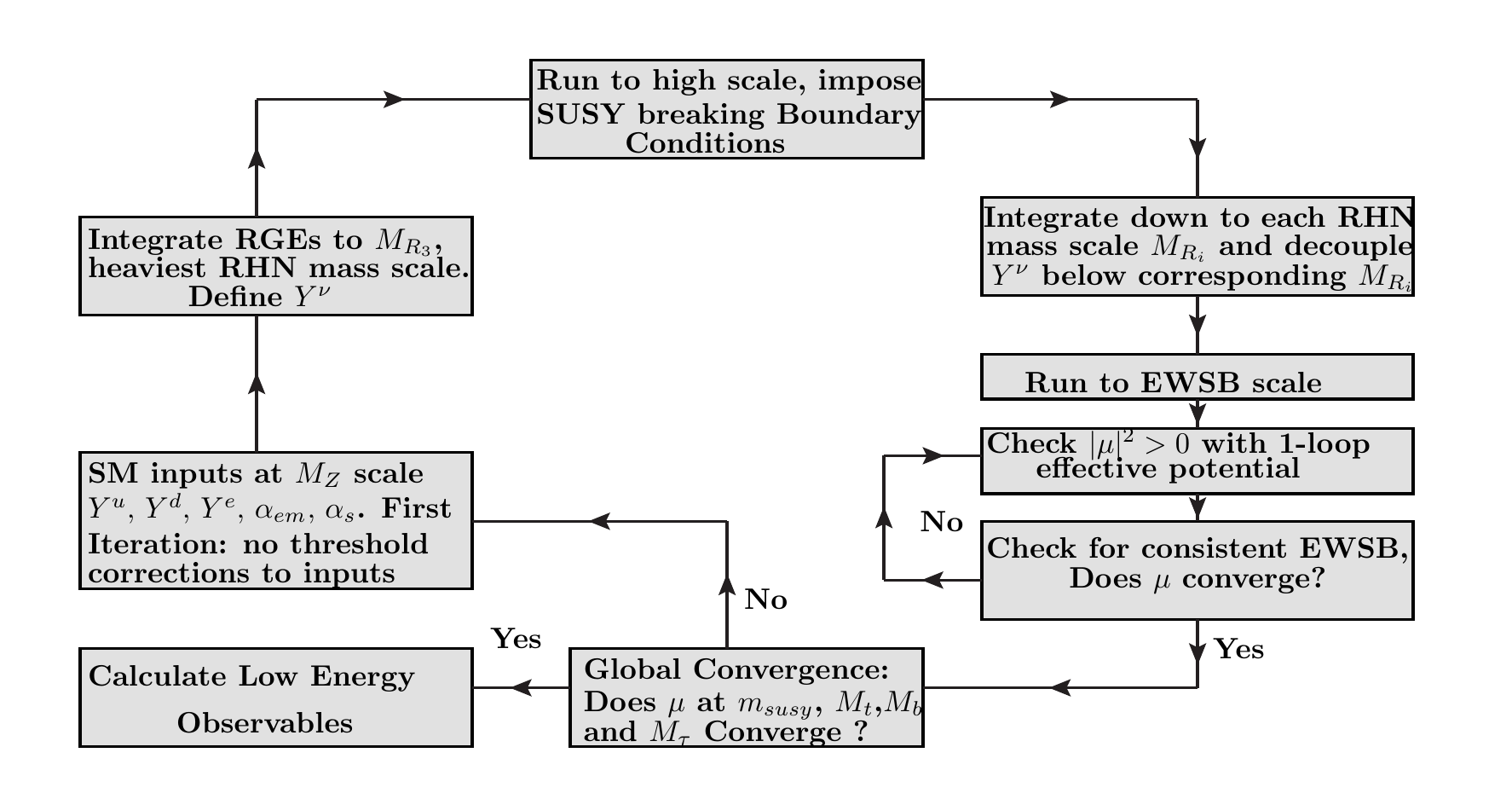}
\caption{Flowchart of \SUSEFLAV ~program}
\end{center}
\end{figure}

\textbf{2.} The numerical procedure implemented in the program can be summarized in a pictorial form presented in Fig. 1.  

\begin{itemize}
\item Step1: Low energy inputs $M_{t}^{pole}$, $M_b^{\overline{MS}}(mb)$, $M_{\tau}^{pole}$, $g_{1,2,3}^{\overline{DR}}$ are defined in $\overline{DR}$ scheme. The CKM matrix is also one of the inputs at the weak scale.
 \item Step 2: The Renormalisation Group Equations (RGE) of the Yukawas and the gauge couplings are first evolved from the weak scale to the heaviest right handed neutrino scale~($M_{R_3}$) where neutrino Dirac Yukawa $Y_{\nu}$ is defined. The RGE are then  evolved to GUT scale, \textit{i.e.,} the scale at which $g_1 = g_2$. If the right handed neutrinos 
 are switched off, the program directly checks for the GUT scale. 
 \item Step 2a: The default choices for the $Y_{\nu}$ are CKM like mixing ($Y_{\nu} = V_{CKM} Y_U$), PMNS like mixing  ($Y_{\nu} = U_{PMNS} Y_U$)
 in the up-type Yukawa mass matrix ($Y_U$) at the  seesaw scale. The user is also free to define her choice of the $Y_{\nu}$ either directly
  or by using the R-parameterisation
 of Casas and Ibarra~\cite{casasibarra}. 
\item Step 3: At the GUT scale we impose boundary conditions corresponding to  mSUGRA or Non-Universal Higgs Masses (NUHM)
or Non-Universal Gaugino masses (NUGM)  to all soft terms and run the corresponding RGE
down to  $M_{\text{SUSY}}$, which is chosen to be 1 TeV for the first iteration and the geometric mean of stop masses in the subsequent
iterations. A completely non-universal  (CNUM) set of boundary conditions can also be provided by
the user. This choice is typically useful while studying models with broken flavor symmetries at high scale. 
\item Step 3a: For the GMSB case, the program evolves only up to the messenger scale which is user defined, instead of GUT scale as
in other models.  
\item Step 4: The program checks  the radiative electroweak symmetry breaking conditions  at $M_{\text{SUSY}}$.  Corrections
from  the 1-loop effective potential are also taken into account. They can significantly modify the $\mu$ parameter, which
in turn modifies the mass spectrum. An iterative procedure is employed to find solutions consistent with the mass spectrum
as well as electroweak symmetry breaking conditions. 
\item Step 5:  One-loop supersymmetric threshold  corrections to $M_{t}(M_Z)$,  $M_b(M_Z)$,  $M_{\tau}(M_Z)$,  $g_{1,2,3}(M_Z)$ and $\sin^2\theta_w$ are computed using the particle mass spectrum  at   $M_Z$.  These form the initial values of the Yukawa
and the gauge  couplings to the RGEs at the weak scale. 
\item Step 6: The complete program is then put in an iterative cycle till sparticle masses at $M_{\text{SUSY}}$ and $ M_t(M_Z)$, $M_b(M_Z) $ and $M_{\tau}(M_Z) $ converge to the desired accuracy. 
\item Step 7: Once the spectrum converges we compute the low energy observables. Some of these are  $(g-2)_\mu$, 
and  lepton flavor violating observables like $\mu \to e + \gamma$ etc. 
\end{itemize}

\textbf{3.}~\SUSEFLAV ~requires  ~\LAPACK\ \cite{lapack} library as a prerequisite for compilation.  Running the provided
Makefile would compile the package. This has been tested on some LINUX distributions  and  on MAC OS-X. 
The package produces three executables when compiled, namely {\tt suseflav}, {\tt suseflavslha} and {\tt suseflavscan}, to facilitate three different functionalities for its users.  To compute the spectrum for a single point the usage of executables {\tt suseflav} and {\tt suseflavslha} is recommended. Whereas, to scan the parameter space the usage  of the executable {\tt suseflavscan} can be used.   \SUSEFLAV\ has two input/output modes. These are the {\tt SLHA2} \cite{slha2} I/O interface and non-{\tt SLHA} or traditional \SUSEFLAV\ I/O interface.  Input files starting with {\tt sinputs} should be used to run the program using the traditional \SUSEFLAV\ interface. Each of these files provides an option to run right handed neutrinos with user defined mixing along with algorithm specific inputs such as the precision of computation. Similarly, input files starting with {\tt slha} should be used to run the program using the {\tt SLHA} interface. The following is a list of input files which are provided with the program. All these files can be easily modified
by the user to suit her/his purposes. 

 \begin{enumerate}
  \item {\tt sinputs.in / slha.in}:  These files contain the input parameters to run cMSSM or mSUGRA models.  
 \item {\tt sinputs-nuhm.in / slha-nuhm.in}: These files contains the input parameters to run models with 
 Non-Universal Higgs masses in mSUGRA. 
  \item {\tt sinputs-nugm.in}:  Input file to run models with Non-Universal Gaugino masses in mSUGRA.
 \item {\tt sinputs-gmsb.in / slha-gmsb.in}:  Input files to run Gauge Mediated Supersymmetric Breaking models.
  \item {\tt sinputs-rpar.in}:  Input file contains the parameters to run Casas-Ibarra R parametrization \cite{casasibarra} of neutrino masses and mixing with constrained MSSM boundary conditions. 
 \item {\tt sinputs-cnum.in}: This is the most generic input file for gravity mediated models.
  All the SUSY breaking parameters at the high scale are defined by the user.  This is useful for studying flavor models. 
 \item{\tt sinputs\_scan.in}:  This input file should be used to scan the parameter space of  mSUGRA or NUHM. 
 \end{enumerate}
 \SUSEFLAV~ can be coupled to Dark Matter routines such as \MICROMEGAS, \DARKSUSY\ and \SUPERISO\ to compute the relic density and direct detection rates. The program is free under the GNU Public License and can be downloaded from the following websites:
\begin{itemize}
\item \url{http://cts.iisc.ernet.in/Suseflav/main.html}
\item \url{http://projects.hepforge.org/suseflav/}
\end{itemize}

\bibliographystyle{pramana}

\end{document}